\documentclass[runningheads]{llncs}
\usepackage[T1]{fontenc}
\usepackage{graphicx}
\usepackage{graphicx}
\usepackage{adjustbox}
\usepackage{mathtools}
\usepackage[utf8]{inputenc} % allow utf-8 input
\usepackage[T1]{fontenc}    % use 8-bit T1 fonts
\usepackage{url}            % simple URL typesetting
\usepackage{booktabs}       % professional-quality tables
\usepackage{amsfonts}       % blackboard math symbols
\usepackage{nicefrac}       % compact symbols for 1/2, etc.
\usepackage{microtype}      % microtypography
\usepackage{color}
\usepackage{bm}
\usepackage{bbm} %empty number
\usepackage{bigstrut}
\usepackage{multirow}
\usepackage{amsmath}
\usepackage{colortbl}
\usepackage{mathrsfs}   %huati
\usepackage{footnote}
\usepackage{setspace}
\usepackage{algorithmic}
\usepackage[linesnumbered,ruled]{algorithm2e}
\usepackage{array}
 \usepackage[T1]{fontenc}
 \usepackage{lmodern}
 \usepackage[misc]{ifsym}
\begin{document}
\title{Diffusion Modelling for Super-resolved Spatial Transcriptomics of Brain Cancer}
\titlerunning{Spatial Transcriptomics Prediction via Cross-Modal  Diffusion modelling}

\author{Xiaofei Wang\inst{1} \and
Xingxu Huang\inst{5}\and
Stephen Price\inst{1}\and
Chao Li\inst{1,2,3,4}}

\institute{Department of Clinical Neurosciences,
University of Cambridge, UK \and
Department of Applied Mathematics and Theoretical Physics, University of Cambridge, UK \and
School of Science and Engineering, University of Dundee, UK \and
School of Medicine, University of Dundee, UK \and
Zhejiang Lab, China
}


%
\maketitle              % typeset the header of the contribution
\begin{abstract}
The recent advancement of spatial transcriptomics (ST) allows to characterize spatial gene expression within tissue for discovery research. 
However, current ST platforms suffer from low resolution, hindering in-depth understanding of spatial gene expression. Super-resolution approaches promise to enhance ST maps by integrating histology images with gene expressions of profiled tissue spots. However, current super-resolution methods are limited by restoration uncertainty and mode collapse. Although diffusion models have shown promise in capturing complex interactions between multi-modal conditions, it remains a challenge to integrate histology images and gene expression for super-resolved ST maps.
This paper proposes a cross-modal conditional diffusion model for super-resolving ST maps with the guidance of histology images. 
Specifically, we design a multi-modal disentangling network with cross-modal adaptive modulation to utilize complementary information from histology images and spatial gene expression.
Moreover, we propose a dynamic cross-attention modelling strategy to extract hierarchical cell-to-tissue information from histology images. 
Lastly, we propose a co-expression-based gene-correlation graph network to model the co-expression relationship of multiple genes.
Experiments show that our method outperforms other state-of-the-art methods
in ST super-resolution on three public datasets. Codes are available at \url{https://github.com/XiaofeiWang2018/Diffusion-ST}.

\end{abstract}

\section{Introduction}

%Spatial transcriptomics (ST) has demonstrated enormous potential  for generating  molecular maps of cells within tissues, 
Spatial transcriptomics (ST), the spatial distribution of gene expressions, enables genomic  profiling while preserving structural information within original tissue, promising to understand complex conditions with spatial heterogeneity.
However, popular ST platforms, e.g., Visium \cite{staahl2016visualization}
%,  SLIDE-seqV2 \cite{stickels2021highly} 
and Stereo-seq \cite{chen2022spatiotemporal}, only measure gene expression in tissue spots, and the very low spatial resolution limits their ability to study in-depth gene expression. Novel biotechnology is developed for targeted in-situ sequencing, e.g., IISS \cite{tang2023improved}, to generate higher resolution ST maps. However, these methods are expensive and time-consuming. Also, they are limited by the technical bottleneck of low capture rates and throughput. 

%e.g.,  super-resolution (SR)-based techniques, can be employed to reconstruct such HR ST data in silicon from LR ST and histology image.

Computational approaches promise to enhance the spatial resolution of ST maps and accelerate scientific discovery \cite{zhao2021spatial,zhou2023spatial,wei2023multi}. For instance, Zhao \textit{et al.} \cite{zhao2021spatial} proposed a Bayesian statistical method using spot neighborhood information for enhancing the resolution of ST maps.
Despite encouraging results, this method is purely based on genomics without leveraging the histology information at the tissue and cellular levels available from ST maps. Therefore, the fine-grained ST is challenged by the biological confidence of imputed resolution. 

Histology features observed from tissue sections are enriched with phenotypic structure and morphology information. Previous studies show that image-level histology features are associated with tissue gene expression\cite{schmauch2020deep,ash2021joint,badea2020identifying,wang2023multi}. This intrinsic link establishes the feasibility of super-resolving ST maps using histology features extracted from tissue images as additional guidance combined with profiled tissue spots. A few studies \cite{pang2021leveraging,bergenstraahle2022super,hu2023deciphering} have attempted this direction, e.g., Pang \textit{et al.} \cite{zhao2021spatial} designed a two-stage transformer-based framework where low-resolution (LR) ST is firstly predicted from histology images and then used to infer high-resolution (HR) ST. However, due to the two-stage design, this model fails to establish the link between the multi-modalities of histology images and gene expression maps. Further, Bergenstraahle \textit{et al.} \cite{bergenstraahle2022super} proposed a multi-scale latent generative model to enhance ST resolution by integrating histology images with gene expression. Nevertheless, the unstable optimization process of the proposed generative model may present significant challenges for enhancing ST maps.

Conditional diffusion models are a class of deep generative models that have achieved state-of-the-art performance in natural and medical images \cite{moser2024diffusion,li2022srdiff,dhariwal2021diffusion}. The model incorporates a Markov chain-based diffusion process along with conditional variables, i.e., LR images, to restore HR images. The conditioning mechanisms in the diffusion model enable the possibility of incorporating multi-modal conditional data, promising better super-resolved outputs.
%The stochastic nature of the diffusion model enables the generation of multiple HR images through sampling, enabling inherent uncertainty estimation of super-resolved outputs. 
Additionally, the objective function of diffusion models is a variant of the variational lower bound that yields stable optimization processes. Given these advantages, conditional diffusion models promise to effectively enhance the resolution of ST maps by integrating genomics and histology images.

However, several challenges remain in developing diffusion models to integrate multi-modal information in super-resolving ST maps: 
%current diffusion-based SR methods are mainly based on a single modality. 
1) %Integrating histology images and LR ST maps into diffusion models increases the number of conditions.
Traditional methods regard multi-modal data as multiple diffusion conditions, which are integrated via simple concatenation or disentanglement \cite{mao2023disc}. These methods treat each modality equally and ignore the modulation across modalities, which may not effectively leverage complementary information in histology images and gene expression;
%, resulting in high-redundancy features for SR; 
2) Diffusion models are primarily intended for the generation of a single image, i.e., treating the ST map of each gene separately.
Intrinsic expression associations across multiple genes can result in inconsistent diffusion processes, posing challenges to the efficient learning of relevant features;
3) The complex structure and morphology information at cellular and tissue levels pose challenges to integrating multi-modal data in the conditioning process of the diffusion model.

To address the challenges, we propose a novel cross-modal conditional diffusion model (Diff-ST) for ST super-resolution (SR). To the best of our knowledge, this is the first diffusion-based multi-modal SR method for ST maps. The main contribution of our work is threefold: 
\begin{itemize}
    \item We propose a new backbone, i.e., a multi-modal disentangling network with cross-modal adaptive modulation, for the conditional diffusion model. The novel disentangled representation learning enables the modulation of complementary information from tissue images and ST maps.
    \item We propose a co-expression intensity-based gene-correlation graph (CIGC-Graph) network
        to model the co-expression relationship of multiple genes, enabling jointly reconstructing SR images of multiple genes.
    \item We propose a cross-attention modelling strategy based on curriculum learning mechanisms, which enables extracting hierarchical cell-to-tissue level information from tissue images.
   % \item We introduce a curriculum learning-inspired feature fusing strategy to reduce the impact of varied histology information complexity on model convergence.
\end{itemize}

Our extensive experiments on the three public datasets demonstrate that our method outperforms other state-of-the-art (SOTA) methods.

\section{Methodology}

\begin{figure}[t]
\includegraphics[width=.85\textwidth]{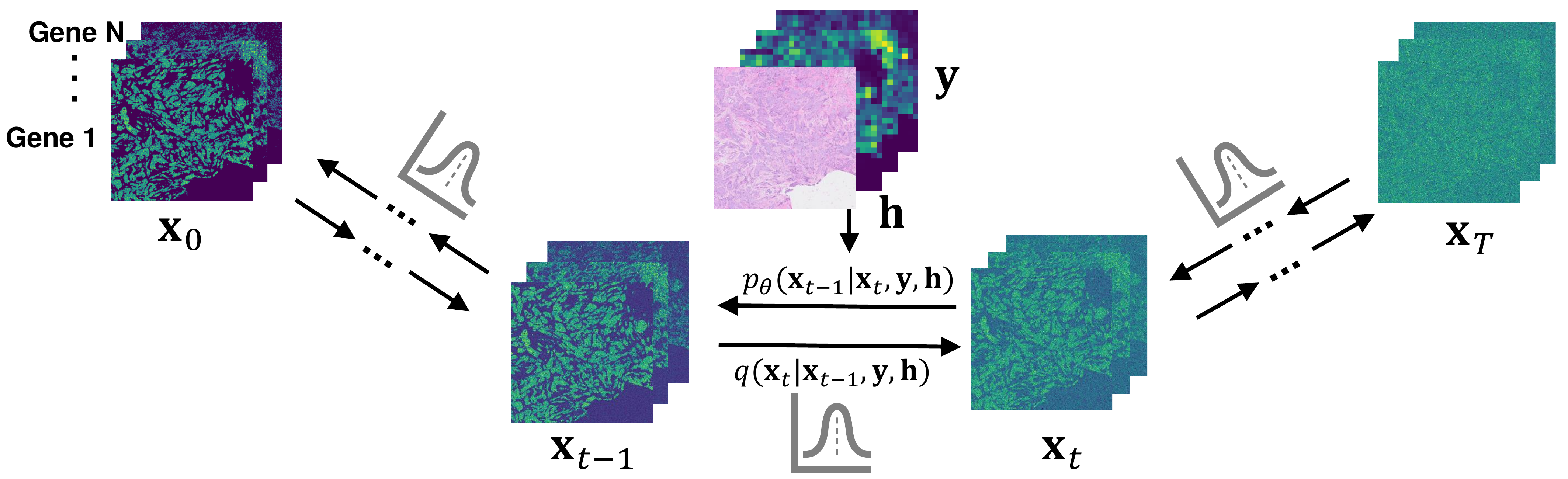}
\centering
\caption{Conceptual workflow of Diff-ST. The forward diffusion process $q$  perturbs HR ST $\mathbf{x}$ by gradually adding Gaussian noise. The backward diffusion process $p$  denoises the perturbed ST, conditioning on its paired LR version $\mathbf{y}$ and histology image  $\mathbf{h}$.}
\label{Fig1}
\end{figure}
\subsection{Framework}
The proposed Diff-ST achieves multi-modal ST SR through forward and backward diffusion processes,  illustrated in Fig. \ref{Fig1}, inspired by \cite{dhariwal2021diffusion}. Given the multi-channel\footnote{Channel number denotes the number of genes that are predicted together.} HR ST images $\mathbf{x}_0 \sim q(\mathbf{x}_0)$, the forward process gradually adds Gaussian noise to $\mathbf{x}_0$ over $T$ diffusion steps according to a noise variance schedule $\beta_1, \ldots, \beta_T$. Specifically, each step of the forward diffusion process produces noisier images $\mathbf{x}_t$ with distribution $q(\mathbf{x}_t \mid \mathbf{x}_{t-1})$, formulated as:
\begin{gather}
q\left(\mathbf{x}_{1: T} \mid \mathbf{x}_{0}\right) = \prod_{t=1}^T q\left(\boldsymbol{x}_t \mid \boldsymbol{x}_{t-1}\right),
    q\left(\mathbf{x}_t \mid \mathbf{x}_{t-1}\right) = \mathcal{N}\left(\mathbf{x}_t ;\sqrt{1-\beta_t} \mathbf{x}_{t-1}, \beta_t \mathbf{I}\right)
\end{gather}

For sufficiently large $T$, the perturbed HR $\mathbf{x}_T$ can be considered as a close approximation of isotropic Gaussian distribution. On the other hand, the reverse diffusion process $p$ aims to generate new HR ST maps from $\mathbf{x}_T$. This is achieved by constructing the reverse distribution $p_\theta\left(\mathbf{x}_{t-1}\mid\mathbf{x}_t, \mathbf{y}, \mathbf{h}\right)$, conditioned on its paired LR ST maps $\mathbf{y}$ and histology image $\mathbf{h}$  as follows: 
\begin{gather}
p_\theta\left(\mathbf{x}_{0: T}\right)=p_\theta\left(\mathbf{x}_T\right) \prod_{t=1}^T p_\theta\left(\mathbf{x}_{t-1} \mid \mathbf{x}_t\right) \nonumber\\
p_\theta\left(\mathbf{x}_{t-1} \mid \mathbf{x}_t, \mathbf{y}, \mathbf{h} \right)=\mathcal{N}\left(\mathbf{x}_{t-1} ; \boldsymbol{\mu}_\theta\left(\mathbf{x}_t, \mathbf{y}, \mathbf{h}, t\right), \sigma^2_t\mathbf{I}\right)
\label{eq2}
\end{gather}
where $p_\theta$ denotes a parameterized model, $\theta$ is its trainable parameters, while $\boldsymbol{\mu}_\theta$ and $\sigma^2_t$ can be learned. Specifically, the detailed conditioning mechanism for reverse diffusion is introduced as follows.

\subsection{Conditioning mechanisms for reverse diffusion}
\begin{figure}[t]
\includegraphics[width=.9\textwidth]{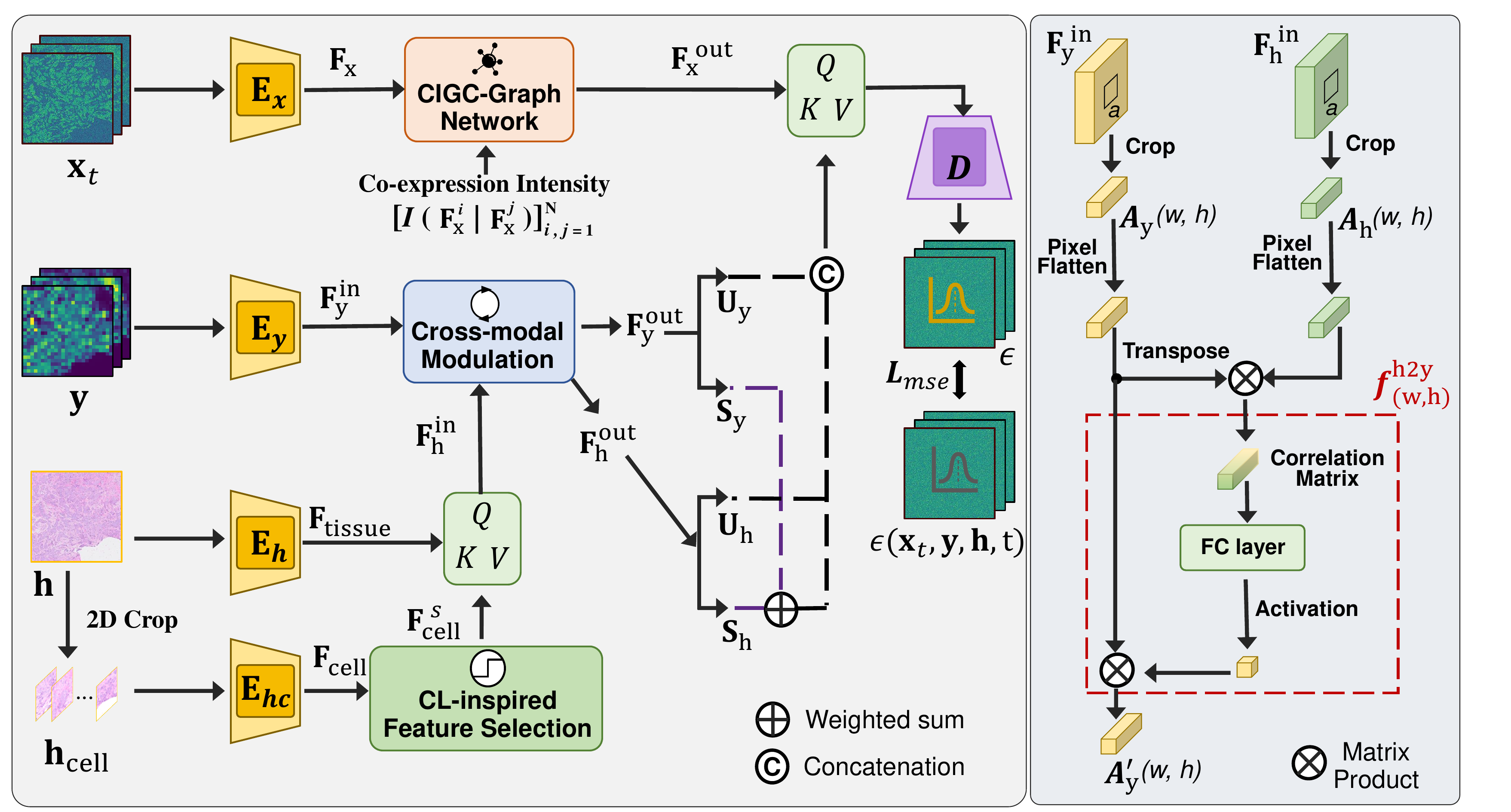}
\centering
\caption{Left: Illustration of the multi-modal conditioned reverse diffusion process of Diff-ST. Right: Pipeline of cross-modal (histology-to-ST) adaptive modulation strategy. CL is curriculum learning, while FC denotes fully connected.}
\label{Fig2}
\end{figure}

\noindent\textbf{1) Overall process:} Taking the multi-modal histology image and LR ST maps as input conditions, our proposed Diff-ST learns to reconstruct the HR ST maps in the reverse diffusion process. Fig. \ref{Fig2}(a) shows the detailed  conditioning process, in which the reverse distribution is estimated by learning multi-modal representations. 
Specifically,  we first extract hierarchical cell-to-tissue level features from histology images via cross-attention modelling.
Then, the extracted features are fused with LR ST features via the cross-modal adaptive modulation and multi-modal disentangling strategy.
Finally, the fused multi-modal features are fed into the reverse diffusion process as conditions for reconstructing HR ST.

\noindent\textbf{2) Hierarchical cell-to-tissue  feature extraction:}
To obtain the cell-level information, we first crop the input histology image $\mathbf{h}$ into $M$\footnote{Here, $M$ is set to 256, with each patch of 320 pixels, reflecting 160$\mu$m regions in real tissue, consistent with the average cellular organization scale in histology \cite{chen2022scaling}.} patches $\{\mathbf{h}_{\mathrm{cell}}\}_1^{M}$ at the cellular scale.
Due to the varied cellular complexity across patches, we propose a dynamic patch selection strategy 
based on curriculum learning. Specifically, given encoded features  $\{\mathbf{F}_{\mathrm{cell}}\}_1^{M}$ of patches, we select and retain the features as $\mathbf{F}_{\mathrm{cell}}^{\mathrm{s}}  =  \{\mathbf{F}_{\mathrm{cell}}^{m}  \big |  \mathcal{E}(\mathbf{h}_{\mathrm{cell}}^{m})>\gamma\}_{m=1}^{M}$, where $\mathcal{E}(\cdot)$ is the entropy-based image complexity estimation function. As training progresses, $\gamma$ gradually increases, indicating increased complexity of 
sampled patches. 
 
 To generate the final histology features, the screened cell-level features $\mathbf{F}_{\mathrm{cell}}^{\mathrm{s}}$ are further integrated with the tissue-level features $\mathbf{F}_{\mathrm{tissue}}$ via the cross-attention modelling
$\text{Atten}(Q, K, V) = \text{softmax}\left(\frac{QK^T}{\sqrt{d}}\right) \cdot V$, with

\begin{equation}
Q = f(\mathbf{F}_{\mathrm{tissue}}) \cdot W_Q , \;
 K =f(\mathbf{F}_{\mathrm{cell}}^{\mathrm{s}}) \cdot W_K , \;
 V = f(\mathbf{F}_{\mathrm{cell}}^{\mathrm{s}}) \cdot W_V . \nonumber
\end{equation}
Here, $f(\cdot)$ denotes the pixel flattening function, and $W_V \in \mathbb{R}^{d_c \times d_t}$, $W_Q \in \mathbb{R}^{d_t \times d} $ \& $W_K \in \mathbb{R}^{d_c \times d}$ are learnable projection matrices, where $d_t$ and $d_c$ denote the features channels of $\mathbf{F}_{\mathrm{tissue}}$ and $\mathbf{F}_{\mathrm{cell}}^{\mathrm{s}}$, respectively.

\noindent\textbf{3) Cross-modal adaptive modulation \& multi-modal disentangling:}
Given the extracted multi-modal features of histology image $\mathbf{F}_{\mathbf{h}}^{\mathrm{in}}$ and LR ST maps $\mathbf{F}_{\mathbf{y}}^{\mathrm{in}}$, we propose a cross-modal adaptive modulation strategy to share the modal-specific information with each other, i.e., cellular and tissue information from histology to ST, and gene expression patterns from ST to histology. Specifically, in the histology-to-ST modulation (Fig \ref{Fig2}(b)), to each pixel in $\mathbf{F}_{\mathbf{y}}^{\mathrm{in}}$ (denoted as $\textbf{y}(w,h)$), we learn a modulation filter $f^{h2y}_{(w,h)}(\cdot)$ constricted to its adjacent pixels in an $a\times a$ region (denoted as $\textbf{A}_{\textbf{y}(w,h)}$) 
and target its counterpart region in $\mathbf{F}_{\mathbf{h}}^{\mathrm{in}}$ (denoted as $\textbf{A}_{\textbf{h}(w,h)}$).
As such, the filter can fully exploit the local dependency between histology images and ST maps. 
The filtering operation is expressed as
\begin{equation}         
  \textbf{A}'_{\textbf{y}(w,h)} = \textbf{A}_{\textbf{y}_{(w,h)}} \textbf{w}^{h2y}_{(w,h)}, 
  \text{where} \; \textbf{w}^{h2y}_{(w,h)} = FC((\textbf{A}_{\textbf{y}_{(w,h)}})^T \textbf{A}_{\textbf{h}(w,h)}) 
  \label{eq_s'}
\end{equation}  
where $FC$ is fully connected layer and the resulting $\textbf{A}'_{\textbf{y}(w,h)}$ is the update of $\textbf{A}_{\textbf{y}_{(w,h)}}$ and is induced to spatially refer to $\textbf{A}_{\textbf{h}(w,h)}$, which is in HR domain.

To further capture the unique and shared representation of multi-modal data, a disentangling network is designed after the cross-modal adaptive modulation. With the modal-unique ($U_\mathbf{h}$ for histology and $U_\mathbf{y}$ for ST) and modal-sharing features ($S_\mathbf{h}$ for histology and $S_\mathbf{y}$ for ST), a cross-modal disentangling loss $\mathcal{L}_\text{cm-dis}= \|S_\mathbf{y} - S_\mathbf{h}\|_2 / \|U_\mathbf{y} - U_\mathbf{h}\|_2$ is designed for minimizing the disparity among shared representations while maximizing that among unique representations.

\subsection{Co-expression intensity-based gene-correlation graph}
In predicting expressions of multiple genes,  existing diffusion models that separately infer multi-gene ST maps may ignore the intrinsic co-expression correlation among genes. 
 We propose a co-expression intensity-based gene-correlation graph (CIGC-Graph) network for effective modelling of co-expression among genes.
   
CIGC-Graph (Supplementary Fig. 1) is defined as $ \mathcal{G}=(\mathbf{V},\mathbf{E})$, where  $\mathbf{V}$  indicates the nodes, while $\mathbf{E}$ represents the edges. Given the encoded features of the noised HR ST maps of $N$ genes
$\mathbf{F}_{\mathbf{x}}=[\mathbf{F}_{\mathbf{x}}^{i}]_{i=1}^{N} \in \mathbb{R}^{N \times C}$ as input nodes, we construct a co-expression intensity based correlation matrix $\mathbf{I} \in \mathbb{R}^{N \times N}$ to reflect the relationships among each node feature, with a weight matrix $\mathbf{W}_{g} \in \mathbb{R}^{C\times C}$ to update the value of $\mathbf{F}_{\mathbf{x}}$. Formally, the output nodes $\mathbf{F}_{\mathbf{x}}^{\mathrm{out}} \in \mathbb{R}^{N\times C}$ are formulated by a single graph convolutional network layer as

\begin{equation}
\label{E1}
\mathbf{F}_{\mathbf{x}}^{\mathrm{mid}} = \mathcal{\delta} (\mathbf{I} \mathbf{F}_{\mathbf{x}} \mathbf{W}_{g}) , \text{where} \; \mathbf{I} = [I_{i}^{j}]_{i,j=1}^{N}, A_{i}^{j} = \frac{1}{2}\big (p (\mathbf{F}_{\mathbf{x}}^{i}|\mathbf{F}_{\mathbf{x}}^{j})+p (\mathbf{F}_{\mathbf{x}}^{j}|\mathbf{F}_{\mathbf{x}}^{i})\big).
\end{equation}
In \eqref{E1}, $\mathcal{\delta} (\cdot)$ is an activation function and $p (\mathbf{F}_{\mathbf{x}}^{i}|\mathbf{F}_{\mathbf{x}}^{j})$ denotes the relative expression intensity of $i$-th gene in terms of $j$-th gene. Besides, residual structure is utilized to generate the final output $\mathbf{F}_{\mathbf{x}}^{\mathrm{out}}$ of CIGC-Graph network, defined as  
$\mathbf{F}_{\mathbf{x}}^{\mathrm{out}} = \alpha \mathbf{F}_{\mathbf{x}}^{\mathrm{mid}} + (1-\alpha) \mathbf{F}_{\mathbf{x}}, $
where $\alpha$ is a graph balancing hyper-parameter.

\section{Experiments \& Results}

\subsection{Datasets \& Implementation Details}
\noindent\textbf{Datasets:}
We evaluated our model on three public datasets, i.e., Xenium \cite{xenium_brest}, SGE \cite{SGE} and Breast-ST \cite{breast_ST}. In all datasets, we use  LR ST maps and paired HR histology images to restore $5\times$ and $10\times$ HR ST maps, aligning with settings in \cite{hu2023deciphering,zhang2024inferring}.
Totally, we include 502 histology images and 12,550 ST maps of 25 genes.

In the Xenium dataset, We randomly split the 232 histology images (with 5,800 ST maps) into 99 (with 2,475 ST maps) for training, 49 (with 1,225 ST maps) for validation, and 84 (with 2,100 ST maps) for testing. For both SR times, the histology images are of 5,120 $\times$ 5,120 pixels at 0.5 $\mu$m $\textrm{px}^{-1}$, while the LR ST maps are of 26 $\times$ 26 pixels at 100 $\mu$m $\textrm{px}^{-1}$. Besides, the HR ST maps are of 256 $\times$ 256 pixels at 10 $\mu$m $\textrm{px}^{-1}$ and of 128 $\times$ 128 pixels at 20 $\mu$m $\textrm{px}^{-1}$ for $10\times$ and $5\times$ SR, respectively.
Moreover,  the SGE (47 histology and 1,175 ST) and Breast-ST (223 histology and 5,575 ST) are with the same resolution setting as Xenium and are both set as external validation datasets.

\noindent\textbf{Implementation details:}
We trained our model for 200 epochs on two NVIDIA RTX A5000 24 GB GPUs, with batch size 4 and learning rate 0.0001 with AdamW optimizer ~\cite{loshchilov2017decoupled} together with the weight decay. 
Following the sampling strategy in \cite{dhariwal2021diffusion}, Diff-ST uses the sample steps of 1,000 in both forward and reverse diffusion processes. Key hyper-parameters are in Table I of supplementary material. All hyper-parameters are tuned to achieve the best performance over the validation set. Our method is implemented on PyTorch with the Python environment.

\begin{table*}[t]
\scriptsize
\centering
\caption{Performance comparisons on three datasets with $5\times$ and $10\times$ enlargement scales. Bold numbers indicate the best results.}
\begin{adjustbox}{}
\begin{tabular}{c||cccc|cccc|cccc}
\toprule
\hline
Dataset          & \multicolumn{4}{c|}{Xenium}  & \multicolumn{4}{c|}{SGE} & \multicolumn{4}{c}{Breast-ST}                                                    \\ \hline
Scale            & \multicolumn{2}{c|}{$5\times$}   & \multicolumn{2}{c|}{$10\times$}  & \multicolumn{2}{c|}{$5\times$}                               & \multicolumn{2}{c|}{$10\times$} & \multicolumn{2}{c|}{$5\times$}          & \multicolumn{2}{c}{$10\times$}                   
\\ \hline 

Metrics          & \multicolumn{1}{c}{RMSE} & \multicolumn{1}{c|}{PCC} & \multicolumn{1}{c}{RMSE} & PCC & \multicolumn{1}{c}{RMSE} & \multicolumn{1}{c|}{PCC} & \multicolumn{1}{c}{RMSE} & PCC 
& \multicolumn{1}{c}{RMSE} & \multicolumn{1}{c|}{PCC} & \multicolumn{1}{c}{RMSE} & PCC\\ \hline \hline

U-Net       & \multicolumn{1}{c}{0.376}     & \multicolumn{1}{c|}{0.198}     & \multicolumn{1}{c}{0.392}     & 0.208    & \multicolumn{1}{c}{0.384}     & \multicolumn{1}{c|}{0.463}     & \multicolumn{1}{c}{0.414}     &  0.510  & \multicolumn{1}{c}{0.356}     & \multicolumn{1}{c|}{0.540}     & \multicolumn{1}{c}{0.401}     & 0.514  \\

U-Net++   & \multicolumn{1}{c}{0.288}     & \multicolumn{1}{c|}{0.256}     & \multicolumn{1}{c}{0.319}     & 0.296    & \multicolumn{1}{c}{0.354}     & \multicolumn{1}{c|}{0.509}     & \multicolumn{1}{c}{0.406}     &  0.538  & \multicolumn{1}{c}{0.303}     & \multicolumn{1}{c|}{0.586}     & \multicolumn{1}{c}{0.358}     & 0.492  \\

AttenU-Net   & \multicolumn{1}{c}{0.396}     & \multicolumn{1}{c|}{0.153}     & \multicolumn{1}{c}{0.459}     & 0.133    & \multicolumn{1}{c}{0.407}     & \multicolumn{1}{c|}{0.413}     & \multicolumn{1}{c}{0.486}     &  0.407  & \multicolumn{1}{c}{0.364}     & \multicolumn{1}{c|}{0.510}     & \multicolumn{1}{c}{0.412}     & 0.477  \\

Guided-DM   & \multicolumn{1}{c}{0.307}     & \multicolumn{1}{c|}{0.263}     & \multicolumn{1}{c}{0.324}     & 0.311    & \multicolumn{1}{c}{0.365}     & \multicolumn{1}{c|}{0.492}     & \multicolumn{1}{c}{0.429}     &  0.483  & \multicolumn{1}{c}{0.276}     & \multicolumn{1}{c|}{0.576}     & \multicolumn{1}{c}{0.279}     & 0.548  \\ \hline

HistoGene   & \multicolumn{1}{c}{0.268}     & \multicolumn{1}{c|}{0.284}     & \multicolumn{1}{c}{0.310}     & 0.321    & \multicolumn{1}{c}{0.346}     & \multicolumn{1}{c|}{0.517}     & \multicolumn{1}{c}{0.412}     &  0.512  & \multicolumn{1}{c}{0.245}     & \multicolumn{1}{c|}{0.610}     & \multicolumn{1}{c}{0.237}     & 0.608  \\

TESLA          & \multicolumn{1}{c}{0.223}     & \multicolumn{1}{c|}{0.328}     & \multicolumn{1}{c}{0.266}     & 0.384    & \multicolumn{1}{c}{0.293}     & \multicolumn{1}{c|}{0.542}     & \multicolumn{1}{c}{0.335}     &  0.550  & \multicolumn{1}{c}{0.197}     & \multicolumn{1}{c|}{0.672}     & \multicolumn{1}{c}{0.208}     & 0.684  \\ \hline 

\textbf{Ours}  & \multicolumn{1}{c}{\textbf{0.120}}     & \multicolumn{1}{c|}{\textbf{0.403}}     & \multicolumn{1}{c}{\textbf{0.173}}     & \textbf{0.471}     & \multicolumn{1}{c}{\textbf{0.171}}     & \multicolumn{1}{c|}{\textbf{0.613}}     & \multicolumn{1}{c}{\textbf{0.195}}     & \textbf{0.626} & \multicolumn{1}{c}{\textbf{0.132}}     & \multicolumn{1}{c|}{\textbf{0.743}}     & \multicolumn{1}{c}{\textbf{0.148}}     & \textbf{0.758}   \\ \hline
\bottomrule
\end{tabular}
\end{adjustbox}
\label{tab1}
\end{table*}

\begin{figure}[ht]
    \centering
\includegraphics[width=.7\textwidth]{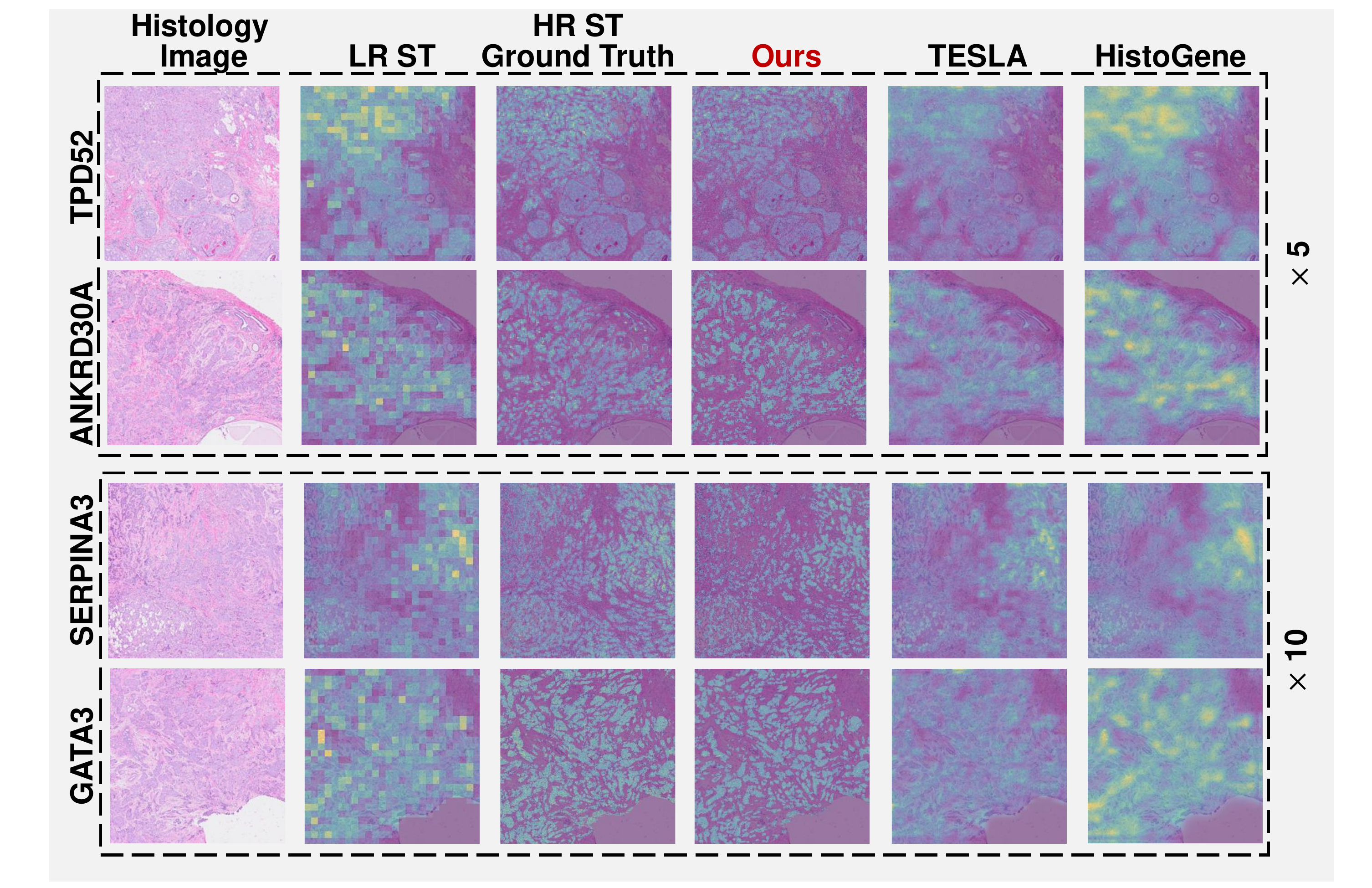}
    \caption{Visual comparisons at $5\times$ and $10\times$ scales on the Xenium dataset. The ST maps are overlayed on the paired histology image for better visualisation. Note that ANKRD30A, TPD52, GATA3 and SERPINA3 denote different genes.  }
    \label{fig:3}
\end{figure}

\subsection{Performance evaluation}
\noindent\textbf{Quantitative comparison on Xenium dataset:} 
We compare our model with six other SOTA methods, i.e., TESLA \cite{hu2023deciphering}, HistoGene \cite{pang2021leveraging},  Guided Diffusion \cite{mao2023disc},  U-Net \cite{ronneberger2015u}, U-Net++ \cite{zhou2018unet++} and  AttenU-Net \cite{oktay1804attention}, at both 5$\times$ and 10$\times$ enlargement scales. Note TESLA  \cite{hu2023deciphering} and HistoGene \cite{pang2021leveraging} are specifically designed for ST SR tasks, while other common image SR methods are baselines. Besides, to ensure the fairness, all the comparison methods use both the HR histology image and LR ST maps for enhancing ST resolution. Experimental results are shown in Table \ref{tab1}, in terms of Root MSE (RMSE) and Pearson correlation coefficient (PCC), consistent with \cite{hu2023deciphering}. As shown, at 5$\times$ scale, Diff-ST performs the best, achieving improvement of at least 0.103 in RMSE and 0.075 in PCC over others, indicating that Diff-ST could effectively integrate histological features and gene expressions for ST SR. Similar results can also be found in 10$\times$ scale. Notably, due to the extremely high heterogeneity in spatial gene expression \cite{miller2021characterizing}, the expression patterns vary greatly across different tissues and genes, making the ST SR task challenging due to complex data distribution and severe class imbalance.

\noindent\textbf{Network generalizability analysis:} We compare our model with SOTA methods on 2 external validation datasets, i.e., SGE and Breast-ST, without fine-tuning. Results are shown in Table\ref{tab1}, where both 5$\times$ and 10$\times$ SR scale settings are tested. We observe that at 10$\times$ scale, our method achieves an RMSE increment of 0.14 and 0.06
 and a PCC increment of 0.076 and 0.074 compared to the best SOTA method, respectively, suggesting our superior robustness and generalizability.

\noindent\textbf{Visual comparison:}
Fig. \ref{fig:3} shows the restoration results of different methods
%our method and the two ST-SR tailored SOTA methods 
at both $5\times$ and $10\times$ scales on Xenium dataset. Diff-ST outperforms all other methods, producing HR ST images with sharper edges and finer details. More visual comparisons are presented in supplementary Fig. 2.

\subsection{Ablation experiments}
  \begin{table*}[t]
  \scriptsize
\centering
\caption{Ablation Study on the Xenium dataset with $5\times$ and $10\times$ enlargement scale. CAM denotes cross-modal adaptive modulation.}
\resizebox{0.7\columnwidth}{!}{%
\begin{tabular}{c||>{\centering\arraybackslash}p{1.2cm} >{\centering\arraybackslash} p{1.2cm}| >{\centering\arraybackslash} p{1.2cm} >{\centering\arraybackslash}p{1.2cm}}
\toprule
\hline
Scale            & \multicolumn{2}{c|}{5$\times$} & \multicolumn{2}{c}{10$\times$}    \\ \hline 
Metrics          & {RMSE} & {PCC} & {RMSE} & PCC \\ \hline \hline
$w/o$  CAM    & {0.186}     &0.388    &  0.203 & 0.434  \\
$w/o$  CIGC-Graph    & 0.172   & {0.384}     &0.195    &  0.428  \\
$w/o$  hierarchical modelling       & 0.164   & {0.377}     &0.213    &  0.456 \\ \hline \hline
\textbf{Diff-ST (Ours)}  & \textbf{0.120}     & {\textbf{0.403}}     & {\textbf{0.173}}     & \textbf{0.471}   \\ \hline
\bottomrule
\end{tabular}
}
\label{table:2}
\end{table*}
We assess the contribution of three key components in Diff-ST: 1) $w/o$ cross-modal adaptive modulation - integrate histology and ST features via feature concatenation; 2) $w/o$ CIGC-Graph - treat each gene of the ST map separately with simple gene channel concatenation; 3) $w/o$ hierarchical modelling - extract only tissue-level features without cell- level image patching. The 5$\times$ and 10$\times$ scale results on the Xenium dataset are in Table \ref{table:2}. All three models perform worse than Diff-ST, suggesting that these components can enhance the overall model performance. The effectiveness of hierarchical modelling demonstrates that hierarchical histology feature extraction can successfully leverage tissue and cellular information for ST restoration.
Moreover, $w/o$ cross-modal adaptive modulation performs the worst, consistent with our hypothesis that traditional conditional diffusion models may not effectively leverage complementary information in multi-modal data of histology and
gene expression.
 
\section{Summary }
 
ST is an edge-cutting biotechnology but is limited by low spatial resolution for in-depth research. This paper presents Diff-ST, a novel multi-modal conditional diffusion model for ST super-resolution. We propose a cross-modal adaptive modulation strategy to model modality interaction for effective integration, 
%Compared with the traditional concatenation-based conditioning mechanisms in diffusion models, the cross-modal disentangled representation learned by our Diff-ST 
which enables the modulation of complementary information from multi-modalities for effective conditioned diffusion modeling.
To effectively extract intra-modal features from histology and gene expressions, we propose a CIGC-graph network to model gene-to-gene relationships alongside a hierarchical histological feature extraction to capture cell-to-tissue relationships.  
Our experiments demonstrate that our model achieves superior and robust performance over other state-of-the-art methods, serving as a potential tool for \textit{in silico} enhancing ST maps, facilitating downstream discovery research and clinical translation.


\bibliographystyle{splncs04}
\bibliography{samplepaper}

\clearpage

\begin{figure}[h]
\centering
\includegraphics[width=0.99\linewidth]{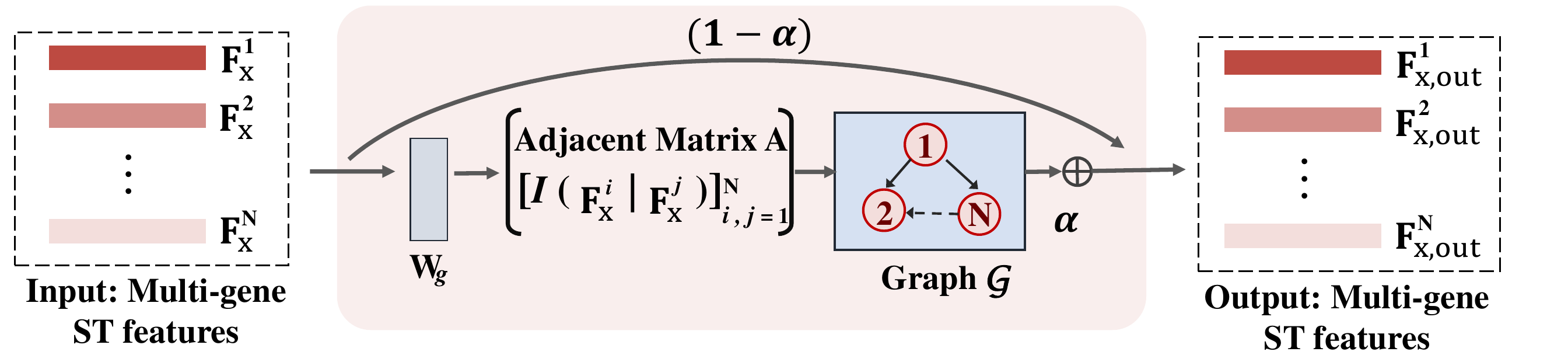}
\caption{ Pipelines of CIGC-Graph network.
}
\label{database}
\end{figure}

\begin{table}[h]
\footnotesize
\caption{Implementation details of our proposed method.}\label{Parameters}
\begin{center}
\begin{tabular}{| l|l|}
 \hline
    Number of genes $N$ for ST maps  & $25$ \\
    Number of cell-level patches $M$ per histology image  & $256$ \\
    Features channels $d_t$ of $\mathbf{F}_{\mathrm{tissue}}$ & $128$ \\
    Features channels $d_c$ of $\mathbf{F}_{\mathrm{cell}}^{\mathrm{s}}$ & $128$ \\
    Intermediate  channels $d$  in cross-attention modelling & $128$ \\
    Region size $a$ in cross-modal adaptive modulation & $32$ \\
    
   Number of features $C$ for the input nodes of CPLC-Graph  & $676$ \\
   Graph balancing weight $\alpha$ &  $0.2$ \\
   
   Exponential decay rate $\beta_{1}$ and $\beta_{2}$ for AdamW optimization &  0.9 and 0.999 \\
   Epsilon $\epsilon$ for AdamW optimization   &  $1\times10^{-8}$\\
   Weight decay  for AdamW optimization   &  $1\times10^{-5}$\\

  \hline

\end{tabular}
\end{center}
\end{table}

\begin{figure}[h]
\centering
\includegraphics[width=0.99\linewidth]{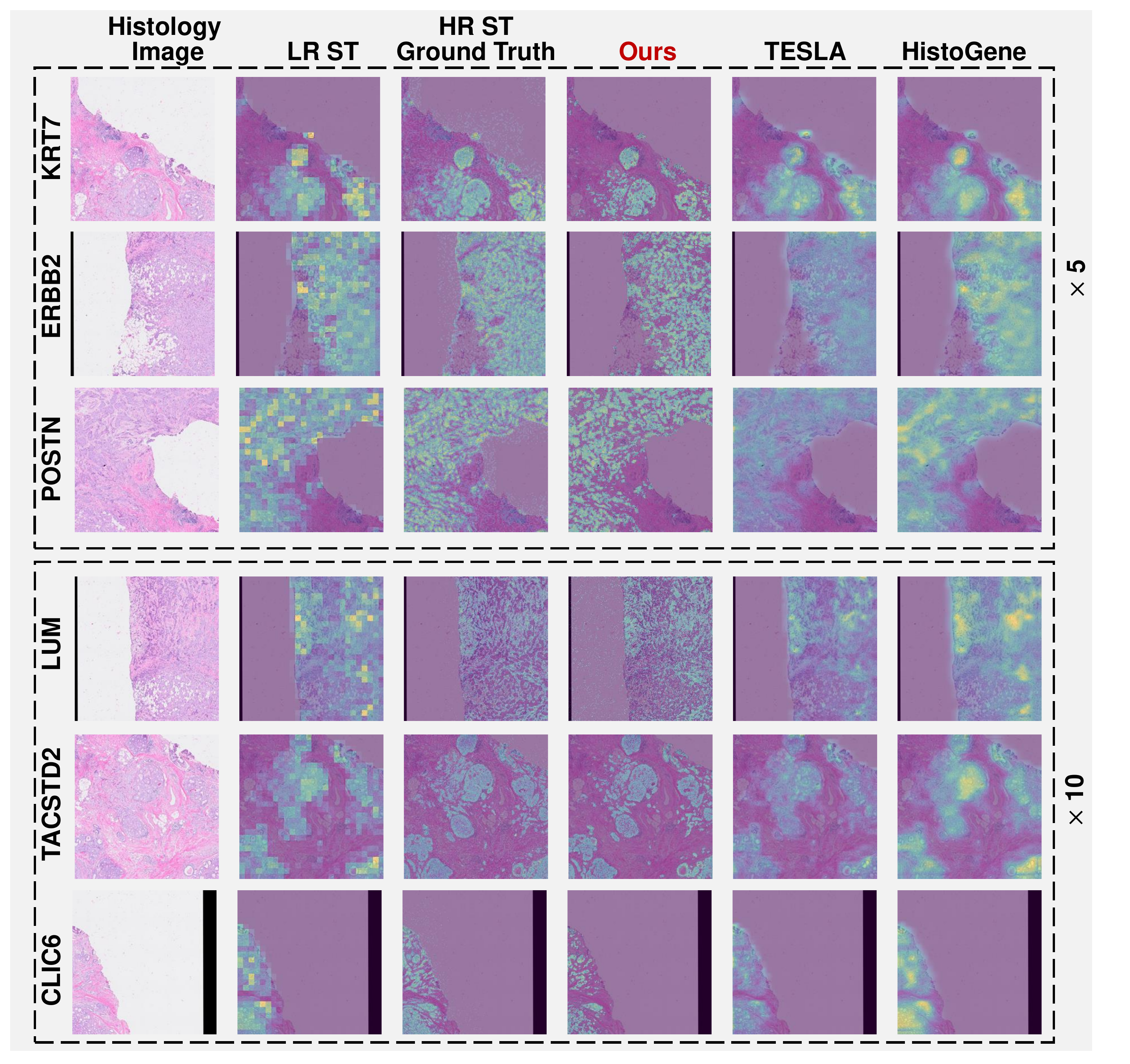}
\caption{ Additional visual comparisons at $5\times$ and $10\times$ scales on the Xenium dataset. The ST maps are overlayed on the paired histology image for better visualisation. Note that KRT7, ERBB2, POSTN, LUM, TACSTD2 and CLIC6 denote different genes.
}
\label{database}
\end{figure}

\end{document}